\documentclass[]{article}
\usepackage{fullpage}
\usepackage{graphicx}

\usepackage{color,url}

\newcommand{\etal}{\emph{et al.~}}
\newcommand{\ie}{\emph{i.e.~}}
\newcommand{\eg}{\emph{e.g.~}}

\newcommand{\anonymous}[2]{#2}

\begin{document}
\title{Simulating the impact of cognitive biases \\on the mobility transition\thanks{Supported by the French National Research Agency (ANR) - SWITCH project}}
\author{Carole Adam}

\date{
Grenoble-Alpes University, Grenoble, France \\
\url{carole.adam@imag.fr}\\
ORCID 0000-0002-2803-8874\\
\url{https://membres-lig.imag.fr/cadam} 
}

\maketitle

\begin{abstract}
Climate change is becoming more visible, and human adaptation is required urgently to prevent greater damage. One particular domain of adaptation concerns daily mobility (work commute), with a significant portion of these trips being done in individual cars. Yet, their impact on pollution, noise, or accidents is well-known. This paper explores various cognitive biases that can explain such lack of adaptation. Our approach is to design simple interactive simulators that users can play with in order to understand biases. The idea is that awareness of such cognitive biases is often a first step towards more rational decision making, even though things are not that simple. This paper reports on three simulators, each focused on a particular factor of resistance. Various scenarios are simulated to demonstrate their explanatory power. These simulators are already available to play online, with the goal to provide users with food for thought about how mobility could evolve in the future. Work is still ongoing to design a user survey to evaluate their impact.
\end{abstract}

\section{Introduction}

In recent decades, the average daily distance traveled by the French population has increased considerably (from 5 km on average in the 1950s to 45 km on average in 2011 \cite{viard2011eloge}), as has the number of personal cars (11,860 million cars in 1970 \cite{barre1997some} compared to 38,3 million in 2021 \cite{ccfa2019,insee}). For example in Toulouse, cars concentrate 74\% of the distances traveled by the inhabitants and contribute up to 88\% to GHG emissions \cite{toulouse}.
The evolution of mobility is therefore an essential question, in the context of the climate crisis but also in terms of public health: the negative impact of a sedentary lifestyle \cite{biswas2015sedentary}, road accidents, air pollution and sound pollution \cite{eea2016}. Indeed, 40000 deaths per year are attributable to exposure to fine particles (PM2.5) and 7000 deaths per year attributable to exposure to nitrogen dioxide (NO2), \ie 7\% and 1\% of the total annual mortality \cite{spf2021}; this report also concludes that the 2-month lockdown of spring 2020 in France made it possible to avoid 2300 deaths by reducing exposure to particles, and 1200 more deaths by reducing exposure to nitrogen dioxide. 
This shows that public policies and individual behaviour changes (modal shift towards cycling, more extensive teleworking) can have an impact on public health. For instance during the COVID-19 pandemics many temporary cycling lanes were set up \cite{rerat2022cycling}. However, aside from such emergencies, such public policies take time to set up, and they are not always well accepted. As an illustration, many of these temporary cycling lanes were returned to cars after the end of the different lockdowns \cite{barbarossa2020post}.

Indeed, despite feeling more and more concerned about climate change, citizens are often reluctant to constraining public policies that could slow it down. Such public policies, including changes in infrastructures or taxes, are not easily accepted (\eg strikes against petrol price rises or new tolls), in particular by those citizens who depend (or believe they do) on their car to commute. As a result, mobilities evolve very slowly, for instance in France a large proportion of commuting is still done by car, even for very short journeys \cite{brutel2021voiture}. Some reasons for this inertia are well-known, such as the lack of alternatives: some rural areas being very poorly served by public transports, while cycling facilities are concentrated in town centres; the cost of electric or newer cars is too high for many workers. Another known reason is the difficulty of changing habits \cite{brette2014reconsidering,lanzini2017shedding}, that are followed routinely unless a life event resets them. Other work show the influence of individualism on user acceptation of radical transition scenarios \cite{epprecht2014anticipating}. Yet another possible explanation of this resistance to mobility change is the influence of cognitive biases in human reasoning \cite{innocenti2013car}.

Cognitive biases are heuristics that facilitate reasoning in situation of uncertainty or danger \cite{tversky1974judgment}. As any heuristics, they are often useful, but can also lead to mistakes with sometimes serious consequences \cite{crash1982}. Research shows that becoming aware of one's biases can help overcoming them in decisions \cite{morewedge2015debiasing}; they thus propose debiasing interventions such as playing a game or watching a video and show their positive impact on short and medium term (2 months). We suggest that interactive simulation could be used as another debiasing intervention; indeed it has been successfully used previously to explain various complex phenomena, for instance the mechanisms of the pandemic \cite{cottineau2020understanding}. 

We therefore propose to use interactive simulators to \textbf{raise awareness about a number of cognitive biases} and their influence on our mobility decisions in the face of climate change. This article describes three simulators designed to illustrate various cognitive biases at work in the resistance to the adoption of so-called soft mobility: habits, reactance bias, and halo bias. These simulators are based on a model of the population as autonomous agents, each agent choosing its mobility according to various criteria, and under the influence of various biases. Each simulator is willingly kept relatively simple, focusing on the role of one particular bias, to facilitate the exploration of links between inputs and outputs. The idea is that people should be able to play with it alone, without needing the guidance or explanations of a supervisor, and still understand or learn something in the process.

This work is part of a larger project aiming at simulating the transition of cities towards more sustainable mobility. Various simulators and serious games have already been proposed in this context \anonymous{\cite{anonymous}}{}. All simulators described in this paper are already available to play online, but have not yet been evaluated. An online survey is being designed for that purpose, and workshops are also being planned in high schools to test the simulators.
The paper is structured as follows. Section~\ref{sec:lit} introduces useful background about cognitive biases.
Section~\ref{sec:habits}, Section~\ref{sec:reac} and Section~\ref{sec:halo} describe our three simulators. Finally, Section~\ref{sec:disc} discusses limitations and prospects of this work, and Section~\ref{sec:cci} concludes the paper.


\section{Background} \label{sec:lit}

\subsection{Cognitive biases}

Cognitive biases are based on heuristics to facilitate decision-making in situations of uncertainty or stress \cite{tversky1974judgment}, and are often useful to make faster decisions despite the lack of information. However, they can also lead to errors with serious consequences. For instance \cite{crash1982} analyse a plane crash as resulting from the pilot's self-deception bias that prevented him from reacting to alerts from his copilot. \cite{murata2015influence} show that it is necessary to recognize and eliminate biases to avoid various accidents or collisions. \cite{doherty2020believing} highlight the importance of educating medical personnel about biases that can affect their decisions and diagnoses. \cite{luz2020heuristics} list various biases that affect patient decisions about vaccination. Finally \cite{mazutis2017sleepwalking} show how biases can explain companies' lack of adaptation to climate change.

\subsection{Some biases that impact mobility}

\cite{innocenti2013car} show that people tend to ``stick'' to the car, even when it is more costly than metro or bus, and they explain this deviation from rational behaviour by the influence of cognitive biases. In this paper, we are particularly interested in three different psychological determinants of ``irrationally sticking'' to the car:
\begin{itemize}
    \item \textbf{Habits}: in line with \cite{brette2014reconsidering} we want to model how habits lead to inertia in the choice of mobility, even when urban infrastructures evolve and are becoming less favorable to driving a car;
    \item \textbf{Reactance} bias, defined as the tendency to react to persuasion attempts that are felt as coercive, by asserting one's free will and strengthening one's non-compliant position as a result. Thus, such messages can have an effect contrary to that intended, by agonising receivers.
    \item \textbf{Halo} bias, defined as the tendency to ignore some negative aspects of an object in order to preserve a previous positive evaluation of it. This bias can lead to ignoring new inconvenients of one's usual mobility mode, for instance an increase in petrol price or traffic jams, in order to avoid a costly questioning of one's mobility.
\end{itemize}

\subsection{Existing models}

Researchers in social simulation for disaster management have previously suggested to consider cognitive biases when modelling human behaviour, for instance during fire evacuation \cite{kinsey2019cognitive} in order to better adapt fire safety buildings design and procedures. There has also been several attempts to model cognitive biases: \cite{arnaud2017role} use BDI (belief-desire-intention) algorithms to model a number of biases preventing people from evacuating when there is a danger of bushfire; \cite{fouillard2021catching} have proposed an analysis of human erroneous decisions in terms of various cognitive biases in the context of plane accidents.

\subsection{Debiasing}
Other researchers therefore propose interventions for "debiasing" individuals and thus help them make the best decisions for themselves and for the common good. \cite{morewedge2015debiasing} prove that becoming aware of our cognitive biases (via interventions such as playing a game or watching a video) can limit their impact on decisions, both immediately and in the medium term (their study spans over 2 months). 

In line with this idea, we propose to use interactive simulators as a debiasing intervention. The idea is to let people freely play with interactive simulators demonstrating the impact of various biases on the mobility choices of a virtual population. 
A similar serious game has been proposed previously to simulate the role of habits in mobility decisions \cite{gamadays22}. This game is implemented with the GAMA simulation platform and cannot be played online. It is also based on realistic geographical data, reproducing an actual medium town from the South-West of France and simulating trips on the actual road network. Besides, it is intended to explain the difficulties of urban planning for the ecological transition, rather than debiasing citizens.

In this paper we propose three simple simulators focused on three different factors of resistance to mobility change: habits (Section~\ref{sec:habits}), reactance bias (Section~\ref{sec:reac}), and halo bias (Section~\ref{sec:halo}). All three simulators are implemented in Netlogo \cite{netlogo99} and available online.

\section{Simulator 1: habits} \label{sec:habits}

Despite the increasing awareness about climate change, and urban planning becoming increasingly favourable to bicycles (cycling lanes, long-term renting services, secured parkings) and unfavourable to cars (traffic jams, difficulty to park, petrol price, pedestrian zones or low-emission zones), the number of people driving to work does not decrease as expected. In this model we show how habits can explain this inertia in the face of evolving urban infrastructures.

\subsection{Conceptual model}

Our model is based on a previous agent-based model of rational mobility choice \anonymous{\cite{anonymous}}{\cite{jacquier2021choice}} but is very simplified. Only 2 mobility modes are considered out of 4 (bicycle and car), with only one criteria of rational choice: how favourable the town infrastructure is to this mode. It is considered here that favouring one mode is necessarily at the expense of the other. The idea is that dedicating more space to bicycles can only be done by taking some space away from cars (transforming car parking slots or car lanes into bike lanes, for instance) and vice-versa. As a result we represent the town infrastructure with a continuous value between 0 (entirely favourable to cars) and 100 (entirely favourable to bicycles).

Citizens are modelled as agents with a very simple behaviour: at every step, they choose a \textbf{mobility} mode and move around with it. They store a \textbf{history} of the mobility mode used each day over their past 20 trips (this value is configurable in the code). This sliding window will be used to compute their habits, once it has reached the minimal length; before that it is considered that the citizen has no habit yet. 

Citizens can choose their mobility in 2 ways:
\begin{itemize}
    \item The \textbf{rational} decision consists in marking each mode based on how favoured it is by the current infrastructure, and then using this mark as a probability to use this mode. This ensures some variability in choices even when one mode is favoured over the other. Initially, the citizens can only choose rationally, while building a history long enough to allow the inference of habits.
    \item The \textbf{routine} decision consists in counting the frequency of each mode over the past 20 trips and using this frequency as a probability to reuse this mode again for the next trip. Over time, the agent's habits might strengthen, giving them more probability to decide routinely rather than rationally.
\end{itemize}

Citizens also have a level of \textbf{satisfaction} or happiness, computed as the rational evaluation of their current mobility mode. We expect that rational deciders switch mode when they become dissatisfied, so they should stay mostly satisfied. On the contrary, routine deciders might keep their usual mode even after it is not satisfying anymore, and so should grow very dissatisfied before they manage to reset their habit.

\subsection{Simulator interface}

The simulator\footnote{Available at: \anonymous{LINK ANONYMISED FOR REVIEW}{\url{https://nausikaa.net/wp-content/uploads/2023/01/habits-mobility.html}}} allows the user to interact with the model in different ways:
\begin{itemize}
    \item A continuous slider can be used to adjust the urban planning, in order to favour cars (value 0), bicycles (value 100), or any intermediate setting. 
    \item A switch controls the use of habits in decisions: if disabled, citizens choose their mode rationally only; if enabled they start building habits that will eventually influence their decisions (as explained above).
    \item A button can suddenly reset all habits, forcing citizens to choose rationally again. This simulates the impact of a crisis (such as the COVID-19 lockdown, forcing everybody to build new habits).
\end{itemize}

\begin{figure}[hbt]
    \centering
    \includegraphics[scale=0.3]{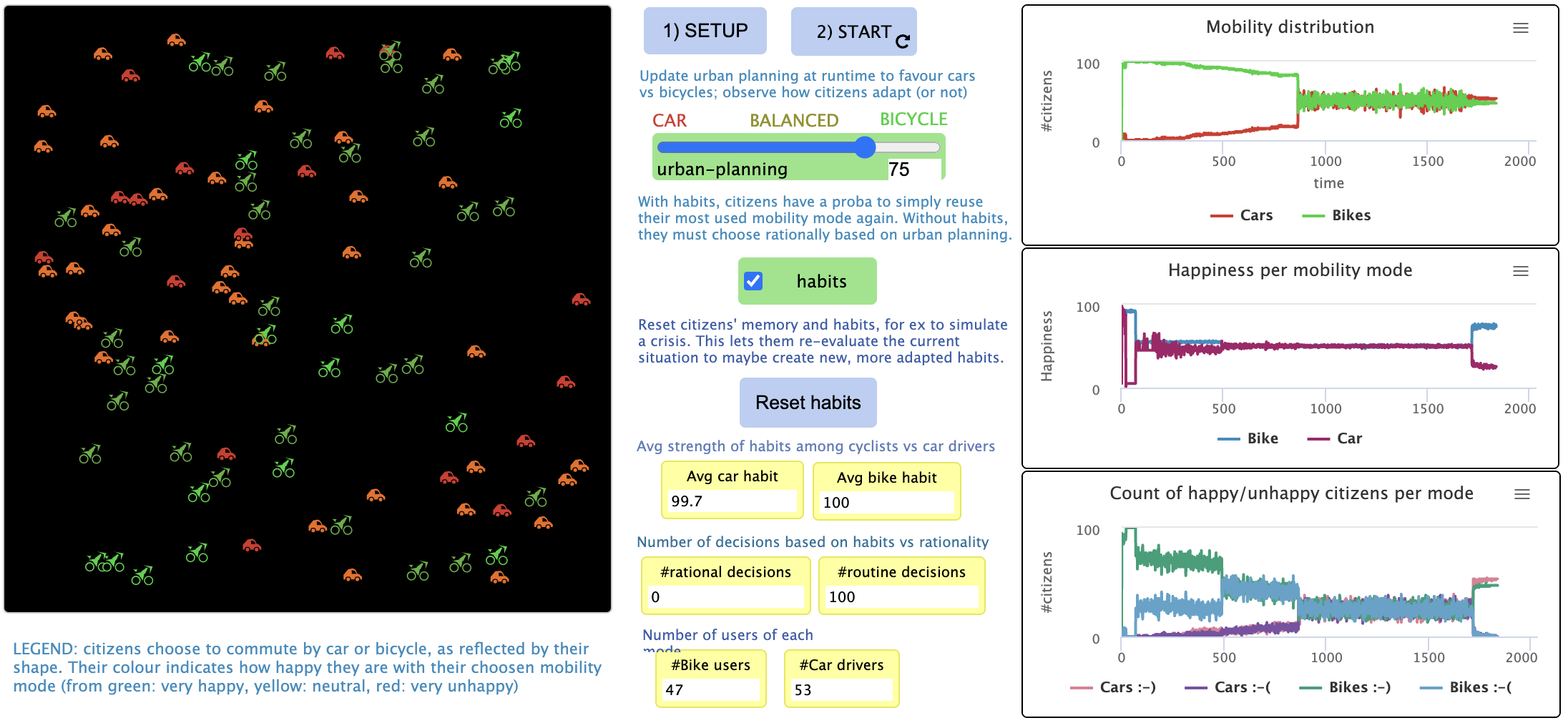}
    \caption{Habits simulator interface}
    \label{fig:habits}
\end{figure}

\noindent The simulator interface (see Figure~\ref{fig:habits} provides various feedbacks for these actions:
\begin{itemize}
    \item Citizens are shown moving around in a window, with their shape (bicycle or car) indicating their mobility mode, and their colour (gradient from red to green) indicating their level of satisfaction (red is very unsatisfied, green is totally satisfied)
    \item The top plot shows the distribution of citizens over the 2 available mobility modes: bicycles (green line) and cars (red line).
    \item The middle plot shows the mean happiness (between 0 and 100) over users of each mode.
    \item The bottom plot shows the count of happy (satisfaction level > 50) and unhappy (satisfaction < 50) users of each mode. 
    \item Monitors display numerical values of: average strength of bike habit and car habit (among their users); number of agents deciding rationally vs routinely; number of users of each mode.
\end{itemize}

\subsection{Results}

\paragraph{Baseline scenario.}
We first ran a baseline scenario without habits, in a town initially very favourable to cars (urban-planning slider set to 10). As a result, most agents use the car and are very satisfied with it (average satisfaction of car users is at around 80\%). We then progressively made the urban planning more favourable to bicycles (final value of 85). The percentage of car users dropped dramatically from around 95\% to only 5\%, while the average satisfaction of the remaining users stayed high (around 70\%). This confirms that without habits, the modal distribution of users reflects the current urban planning, and their average satisfaction is good whatever their mode (otherwise they would switch to the other mode).

\paragraph{Scenario 2: building habits.} We started with the same urban setting very favourable to cars, but this time we activated habits. As we change urban setting to favour bicycles, the number of car drivers decreases, but does not drop as fast as before: habits create an inertia in the citizens' reaction to the evolving urban planning. Besides, the average satisfaction of car drivers drops with the new infrastructure. Indeed, their strong habit pushes them to keep choosing the car, even though they are now clearly dissatisfied with it. This is also visible on the last graph, where most users of the bicycle are happy, while most users of the car are unhappy.

\paragraph{Scenario 3: time of crisis.} Starting from the final situation of the previous scenario, we press the "reset habits" button. This suddenly erases the history of past trips for all citizens, as well as the resulting habits. As a result, all citizens are forced to make a rational evaluation of the available modes again. Since the infrastructure is now clearly favourable to cycling, this will reflect in their mobility choice. Indeed, most agents immediately switch to bicycle, and their happiness grows as a result. They then start building a new habit, pushing them to keep using their bicycle in the future. This scenario shows how times of crisis, such as the COVID-19 pandemic and associated lockdowns, are an opportunity to reset old habits and build new ones, that might be better.

\subsection{Discussion}

This model is willingly very simplified with the idea that it makes it easier to understand its outputs. In particular, citizens are moving around randomly, there are no buildings or roads; indeed, itinerary planning is not relevant here, we are only interested in the impact of habits on modal choice. Besides, such habits should be linked with a specific context (time of day, weather, activity justifying the trip...). This model could also be extended, for instance by linking growing dissatisfaction to an increasing probability to reconsider one's habits. However, we believe that this level of detail would be superfluous to visualise the inertia created by habits, and would only add useless complexity. The goal of this model is simply to illustrate the strength of habits, and how times of crisis that reset them can be exploited to set new, and better, habits for the future.

\section{Simulator 2: reactance bias} \label{sec:reac}

Reactance is defined as a tendency to react inversely to persuasion attempts when they are felt as coercive. The individual might thus strengthen their non-compliant opinion or intention in order to assert their undangered free-will. Persuasive messages can therefore have an effect contrary to that intended, due to agonizing their receivers. We claim that this bias plays a role in reacting to communication campaigns in favour of soft mobility or ecology: part of the population might get angry at these attempts to force them into a constraining behaviour, and react by asserting their right to behave as they want.

\subsection{Conceptual model}

\paragraph{Opinion diffusion model. }
Our conceptual model is interested in how agents react to an official message. We consider opinions about one single fact, modelled as a real number on a continuum from 0 to 1 representing the degree of agreement about that fact. Each agent has its own opinion about this fact (for instance "it is necessary to commute by bike rather than car"), that differs from the other agents' opinions.

When 2 agents meet, they ''share'' their opinions, and as a result, each agent tends slightly towards the opinion of the other. Concretely, we compute the new opinion as an asymmetrical average where one's own opinion weighs more (the exact weight value is configurable in the code) than the received opinion. With this opinion diffusion model, if letting the agents meet each other for a while, the resulting opinions will tend towards an average or compromise, in the middle of the continuum.

\paragraph{Confirmation bias. }

In a first model, agents can be endowed with a confirmation bias. In that case, when meeting another agent, they will only adjust their opinion if this agent agrees with them; on the contrary they will discard opinions that do not confirm theirs. As a result, opinions can only be reinforced, and will this time tend towards the extremes. After a number of interactions, two groups of opposite opinions emerge, the population is polarised.

\paragraph{Reactance bias. }

Based on this first model of opinion diffusion, we have designed a new model where an official messenger (a different type of agent) braodcasts a persuasive message to the population. This message could be for instance a campaign in favour of ecology, soft mobility, or physical activity for health. When agents meet the messenger, they are confronted to its opinion.

Additionally, agents can be susceptible or not to the reactance bias. Non-susceptible agents will get gradually persuaded by the message, as above. Susceptible agents might trigger the reactance bias based on a distance condition: when the agent is exposed to another opinion, it evaluates its distance with its own current opinion. If this distance is small enough, the new opinion is acceptable, and the agent will tend towards it (asymmetrical weighed average, as above). On the contrary, if this distance is ``too far'', \ie over a certain threshold (configurable), the agent will activate the reactance bias. In this case the agent adjusts its opinion away from the official message, reinforcing its disagreement even more.

\paragraph{Message targets. }
Based on this model, for a given message content and a given reactance triggering threshold, agents can fall in one of three different categories:
\begin{itemize}
    \item Already convinced agents, whose opinion match the content of the message;
    \item In-target agents, or positive target: they are not convinced yet, but can still be. Either they are not susceptible to the reactance bias, or they are but the message falls close enough to their own opinion to be acceptable and not trigger the bias. 
    \item Off-target agents, or negative target: they can never be convinced with the current message, because they are susceptible to the reactance bias and their current opinion is far enough to trigger it. As a result, they will react in the opposite direction.
\end{itemize}

\subsection{Simulator}

\begin{figure}[hbt]
    \centering
    \includegraphics[scale=0.3]{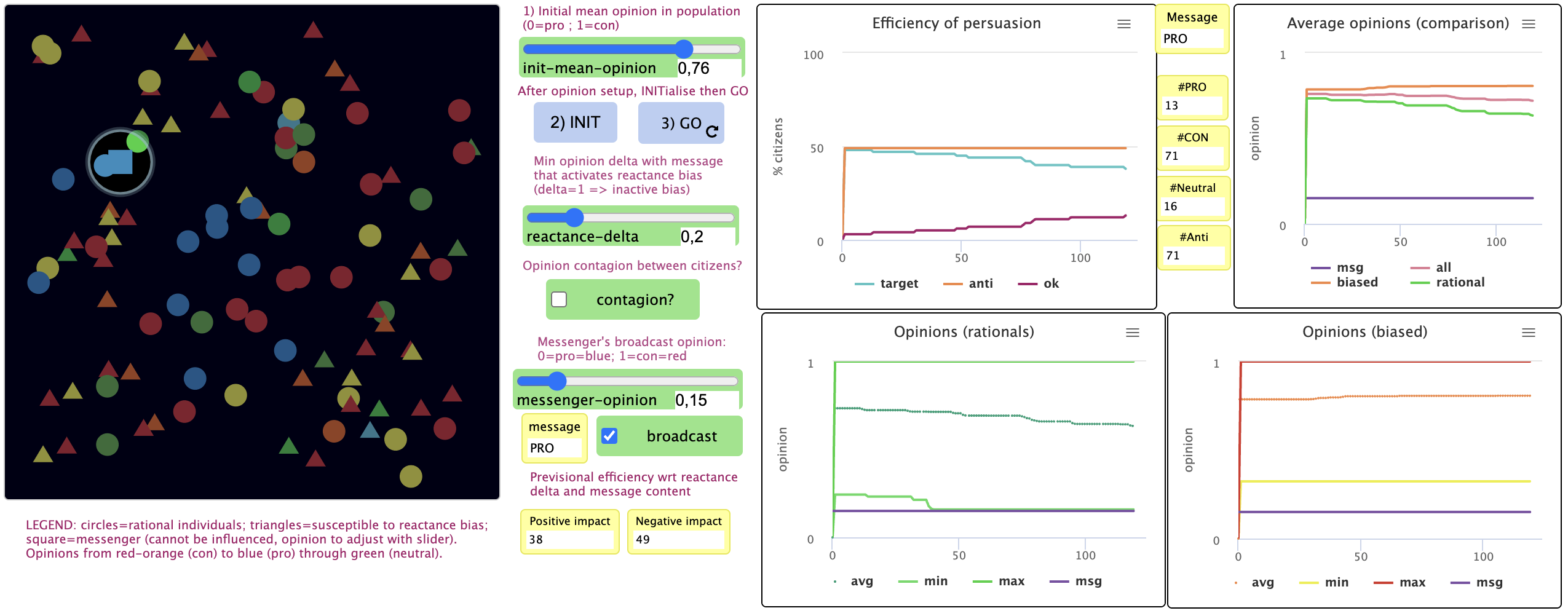}
    \caption{Interface of the reactance bias simulator}
    \label{fig:reac}
\end{figure}

The simulator\footnote{Available at: \anonymous{LINK ANONYMISED FOR REVIEW}{\url{https://nausikaa.net/wp-content/uploads/2023/02/reactance-v2-en.html}}} was implemented in Netlogo. Its interface is shown in Figure~\ref{fig:reac}.

\paragraph{Population initialisation}

The user can choose the desired initial average opinion in the population on the continuum from 0 to 1, by using a slider. When pressing the ``INIT'' button, the population is created with opinions distributed on a Gaussian centered around that mean, and a variance of 0.25. Opinions are kept bounded between 0 and 1 at all times. Each agent's opinion is visually represented by a colour on a gradient between blue (value 0) and red (value 1), with undecise agents ranging from yellow to green. 

Each agent has another attribute determining if it is sensitive to the reactance bias. This is visualised by their shape: circle agents are ``rational'', \ie they do not have the reactance bias; triangle agents are susceptible to the bias, \ie they might activate it when the conditions are fulfilled (message too far from what they can accept).
Once the population is created, the user can start the simulation with the ``GO'' button: the agents then start moving around the window randomly, and can meet other agents or the messenger (represented by a square). The simulation stops automatically when the boradcast message has no impact anymore (\ie its positive target is empty).

\paragraph{Input parameters}

At runtime, the user can modify a number of parameters, regarding the message, and regarding the mechanics of opinion diffusion. First, they can use a slider to choose the value (on the same opinion continuum) of the official message that they wish to broadcast, and use a switch to suspend or resume broadcast. The messenger (shown as a square) will be displayed with the colour matching the content of its message (same colour gradient) as long as it is broadcasting, and will hide when the broadcast is suspended. 

The mechanics of opinion diffusion can be influenced in two ways. First, a contagion switch allows to decide if standard agents (other than the messenger) also share their opinion with each other. If enabled, agents will be influenced both by the messenger and by their neighbours. If disabled, only the messenger can influence the agents' opinions. Second, a slider allows to configure the opinion delta that triggers the reactance bias (for those agents that are susceptible, \ie the triangles). Concretely, if the value is set to 1, it means that only opinions distant by more than 1 from one's own opinion will trigger the bias, so it is in fact disabled. The smaller the value, the easier it is to activate the bias, and the harder it is for the messenger to persuade without agonising its target.

\paragraph{Outputs and feedback}

The interface displays useful information in various forms:
\begin{itemize}
    \item When the user selects the opinion delta triggering the bias, and the content of the message, two monitor boxes update in real time the resulting size of the targets: the positive target are those agents who can be persuaded, while the negative target is made up of those agents who will react reversely due to the reactance bias. The user should try to minimise the size of the negative target so that the message does not have the inverse effect. When the size of the positive target reaches 0, the message has no impact anymore and the simulation stops.
    \item The top left graph illustrates the efficiency of the current persuasive message. The blue line is the percentage of agents that are in the target; it decreases with time as these agents become convinced. The pink line is the percentage of convinced agents; it increases with time as the messenger persuades more agents. Finally the orange line is the percentage of agents that cannot be convinced; it does not change as long as the user does not modify either the message content or the reactance threshold. 
    \item The top right graph compares the official message (purple line) with the average opinion of three groups: the total population (pink line), the rational population not susceptible to reactance (green line), and the biased population (orange line). 
    \item At the bottom, two graphs further detail the opinion dynamics in the rational population (left) and the biased population (right), by showing the minimal and maximal opinion in addition to the average. This allows to see the full range of opinions in each group, and how homogeneous they are. If the persuasion is efficient, these lines should converge towards the message. 
\end{itemize}

\subsection{Scenarios}

We have run a number of scenarios to demonstrate the explanatory power of our simulator.

\begin{figure}[hbt]
    \centering
    \includegraphics[scale=0.3]{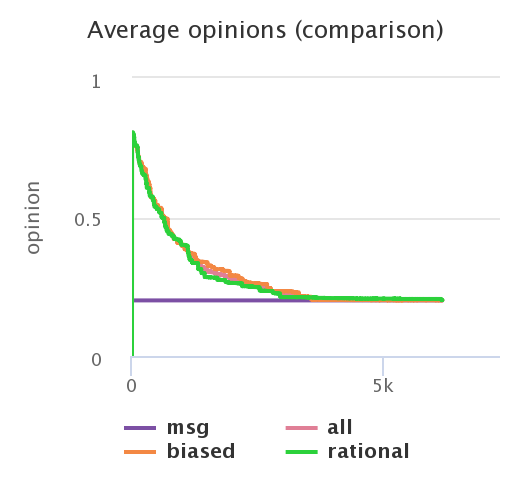}
    \caption{Average opinion dynamics without reactance bias: convergence}
    \label{fig:reac1}
\end{figure}

\paragraph{Scenario 1: baseline. }
We initialised the population with an average opinion of 0.8, and the message with a content at 0.2, so quite distant. For this baseline scenario we set the reactance delta at 1, so that no agent can trigger the bias. We toggle off the opinion contagion between agents in order to isolate the effect of the messenger. We then start broadcast and let the messenger move around and try to persuade citizens. The opinion dynamics graph (see Figure~\ref{fig:reac1}) shows that all average opinions gradually converge towards the official message.

\begin{figure}[hbt]
    \centering
    \includegraphics[scale=0.3]{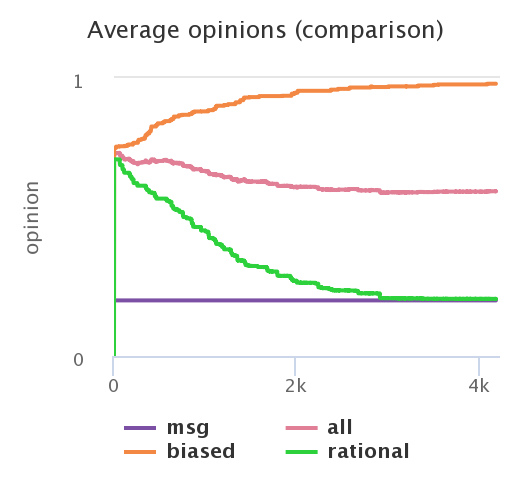}
    \caption{Average opinion dynamics with reactance bias: divergence}
    \label{fig:reac0}
\end{figure}

\paragraph{Scenario 2: reactance. } 
We initialised the population and messenger with the same values as above, and still toggled contagion off, but now set the reactance delta to 0, so that all susceptible agents trigger the bias. As a result, we observe that the two groups (rational vs biased) react differently, as shown by the average opinion graph (Figure~\ref{fig:reac0}). The average rational opinion converges to the content of the message, while the average biased opinion diverges away from the message. 

\begin{figure}[hbt]
    \centering
    \includegraphics[scale=0.3]{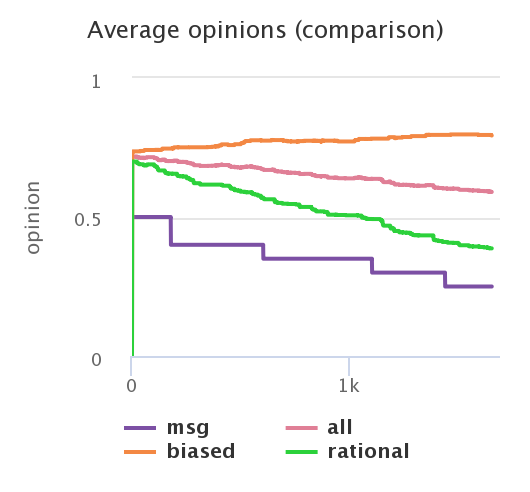}
    \caption{Progressively adapting the content of the message to avoid triggering reactance}
    \label{fig:reacsteps}
\end{figure}

\paragraph{Scenario 3: efficient persuasion. }
We now set the initial opinion to 0.75 and the reactance delta to 0.25 so that some agents can trigger the bias while others will not. The idea here is that the messenger will try to progressively bring the average opinion to 0.25, but in order not to agonize citizens who disagree, it will start with a softer message of 0.5 and progressively change the content towards 0.25. At 0.5, some biased agents are close enough to let themselves be persuaded, while the most extreme opinions will become even more extreme. As the average opinion becomes close enough to 0.5, the messenger can decrease the content of the message to 0.4, then 0.3 and finally 0.25. As shown on Figure~\ref{fig:reacsteps}, the average opinion diverges less since only some agents trigger the bias while some others are drawn towards the official message.

This is in line with classical persuasion strategies of ``putting a foot in the door'' or inciting people to doing small gestures rather than requesting a big change that is more likely to discourage or agonize them. It applies for instance to energy saving gestures to fight global warming (switch your devices off rather than in standby); mobility change (take your bicycle once a week); or resuming physical activity (take the stairs instead of the lift, get out of the bus one stop early to walk, etc).

\paragraph{Scenario 4: inter-individual contagion. }
In the first scenarios, there was no contagion between citizens, so that only the messenger would influence other citizens. If we now toggle contagion on, the result is much more variability in opinions, since citizens can be influenced by their neighbours (whatever their opinion) while the messenger is away in another part of the window. If there are many biased agents with very strong opposite opinions, they might cancel its message faster than it can spread it.

This is similar to the effect of social networks, where people can always connect to others who also disagree with the ``official'' opinion, thus reinforcing their own doubts or opposition, until they might fall in a spiral of conspiracy theories. It forces the authorities to find efficient communication strategies to fight fake news and disinformation.

\subsection{Discussion}
In conclusion, this simulator shows that the further a broadcast message is from their current beliefs, the more individuals risk activating a reactance bias that consists in asserting their free will by adopting the action contrary to the recommendations (\eg ``I will commute by car if I want''). The recommendations must be more progressive so as not to antagonize citizens. This is also the case with many campaigns that recommend ``small gestures''. The idea is that it is dangerous to lose citizens who will build increasing distrust and resentment towards official communication, which will affect all further interactions. The way we modelled reactance as triggered by a distance between opinions could be debated. In particular, another idea would be to model reactance to repetition: the more a message is repeated (echo chamber) the more it might trigger adverse reactions from people who get tired of it. 

The simulator also shows the difficulty of fighting against online disinformation and fake news (when activating the opinion contagion option). It could be gamified by allowing the user to create several messengers, counting the total time and resources invested, and matching it to the results obtained in terms of percentage of the population actually convinced.

\section{Simulator 3: halo bias} \label{sec:halo}

Our third simulator deals with the halo bias, which consists in ignoring certain negative aspects in the evaluation of an object, when they are inconsistent with a positive first impression. This bias thus allows to preserve that position impression. With respect to mobility, it could thus lead to ignoring new inconvenients of one's usual mobility mode, for instance an increase in petrol price or traffic jams, in order to avoid a costly questioning of one's habits.

\subsection{Conceptual model}

\paragraph{Rational multicriteria decision model}
The model is based on an existing model of rational multi-criteria evaluation of modes of mobility \anonymous{\cite{anonymous}}{\cite{jacquier2021choice}} in which 4 modes of travel (car, bike, bus, walking) are evaluated on the basis of 6 criteria (time, cost, comfort, safety, ecology, praticity) and their priority for each individual. Concretely each individual is defined with their personal priorities for the 6 criteria (for instance one individual can be very focussed on ecology but not care about comfort so much, while another will have a big priority on time but not focus on price too much). These priorities are supposed to be static.

Each mode has a mark on each criteria depending on the context (town setting, weather, etc). For instance bicycle and walking are always very ecological while car is not at all; their safety depend on cycling lanes or speed limit; their comfort depend on the temperature while the car or bus are always more comfortable; etc. These marks can evolve with the evolution of the environment (new public policies, time of day, weather...).

Each agent then computes the global mark of each available mobility, as a weighed average of the mark times priority of all 6 criteria. This ensures that different agents might rate the same mobilities differently in the same context, due to their different priorities. Agents are fully rational, and choose the mobility mode that has the best global mark. As a result, when the urban planning evolves (for instance new cycling lanes are built, or the petrol price increases), the citizens might adapt their choice of mobility to the new setting.

\paragraph{Halo bias model}
Our model enriches this rational decision model with the halo bias. Concretely, agents have 6 priorities for the 6 criteria, and can be susceptible to the halo bias or not. Agents that are not susceptible will use the rational decision algorithm as described above. Agents who are susceptible might trigger the halo bias when re-evaluating their current mobility. The triggering condition is based on the distance between the \textbf{value} of this mobility on a criteria, and the \textbf{priority} of that criteria for the agent. If this distance is bigger than a certain delta, then the agent activates the halo bias and ignores this criteria altogether, by setting its priority to 0 when computing the new mark. This leads agents who have adopted a certain mode of mobility to ignore the criteria on which their current mode is poorly rated, in order to maintain a positive overall view. The idea is that it is less costly to question one's priorities than to change one's mobility mode. The halo bias does not influence the evaluation of the other modes.

Finally, agents also have a level of satisfaction or happiness, computed as the mark given to their current mobility. It is expected that rational agents will switch to a different mobility when they grow unhappy with their current mode, while biased agents will ignore the negative criterion in order to restore satisfaction without having to change mobility.

\subsection{Simulator}

\begin{figure}[hbt]
    \centering
    \includegraphics[scale=0.3]{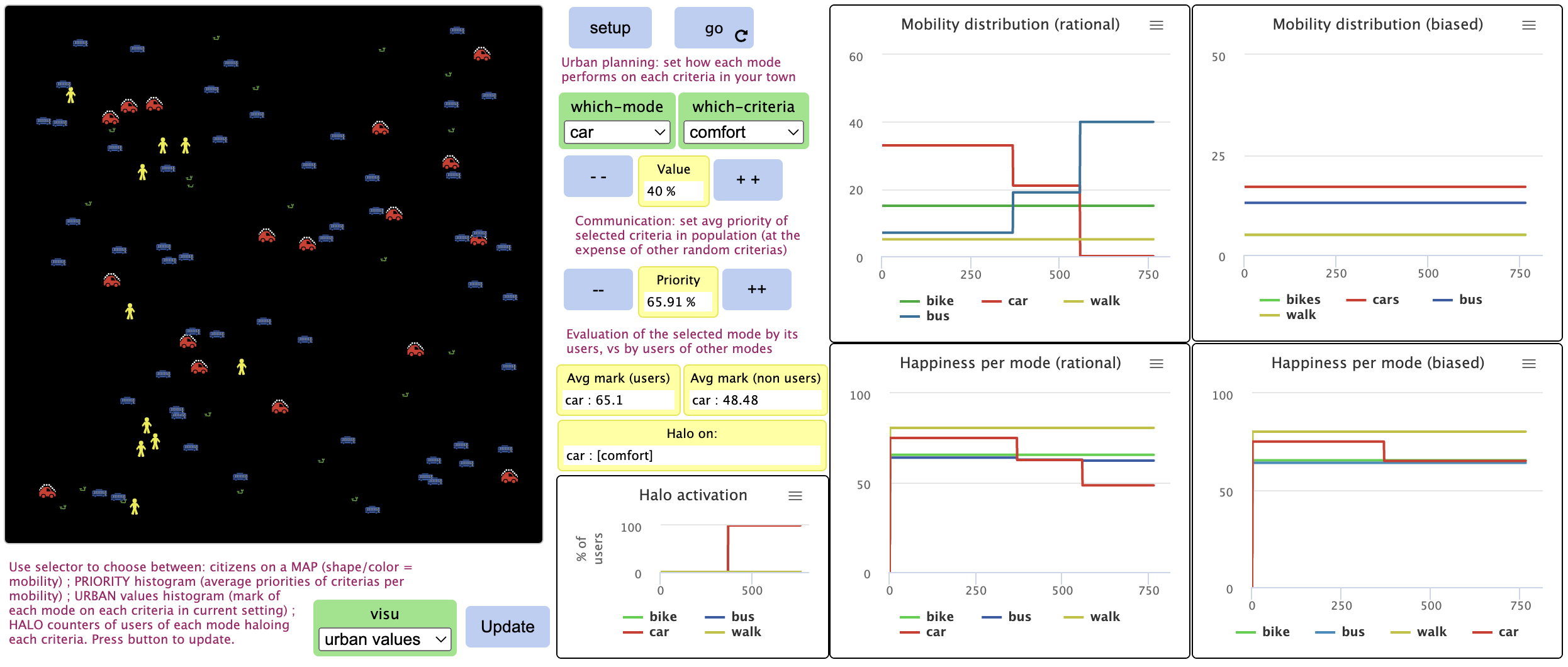}
    \caption{Screeshot of the halo bias simulator interface}
    \label{fig:halo}
\end{figure}

The simulator\footnote{Available at: \anonymous{LINK ANONYMISED FOR REVIEW}{\url{https://nausikaa.net/wp-content/uploads/2023/01/switch-halos-en.html}}} is implemented in Netlogo, and a screenshot of its interface is shown on Figure~\ref{fig:halo}.

\paragraph{Initialisation}
The initial urban planning (scores of the 4 modes on the 6 criteria) is empirically initialised to values matching a standard town: bike and walk are more ecological than bus, while the car is not at all; car and bus are more comfortable (eg less exposed to the weather) than bike or walk; walk is the slowest mode; etc. The exact initial values are visible and configurable in the code. 

The population is then initialised with a believable distribution of citizens on the 4 modes (50\% cars, 20\% bikes, 20\% bus, 10\% walk), and their initial priorities are set so that their rational choice in the current urban setting is indeed their chosen mode, but with some random variability so that not all users of each mode have exactly the same priorities. Again, this is configurable in the code, although our goal to visualise the impact of a bias does not require this distribution to match reality exactly.

Finally, half agents are susceptible to the halo bias, while the other half is not. The triggering threshold is set to 15: this means that when the score of the current mobility mode on a given criterion (rated between 0 and 100\%) falls at least 15 points below the priority of that criterion for the agent (also rated between 0 and 100\%), this agent will activate the halo bias and start ignoring this criterion altogether. The halo bias does not influence the evaluation of the other available modes, which are rated rationally.

\paragraph{User interactions}

The interface allows the user to:
\begin{itemize}
    \item Modify the \textbf{urban planning} of the virtual town, by directly setting the score of each mobility mode on each criterion (what a different urban design would do, such as increasing the frequency of buses or decreasing the maximum authorized speed on roads). All changes are possible in the interface, even though some of them are semantically impossible (\eg no urban policy can make walking not ecological), so it is down to the user to choose wisely.
    \item Modify the \textbf{priority} of each criterion in the population (what a communication or advertising campaign could do). Again, reality is not that simple, no advertising campaign can ensure that ecology suddenly gets maximal priority for eveybody, but this option allows to explore what could happen as a result.
\end{itemize}

\paragraph{Outputs and visualisation}

The interface has three parts. On the left, a dropdown menu below the main window allows the user to choose between four visualisation options. On the main map, the agents are shown with the shape and color indicating their mobility mode (bicycle - green; bus - blue; walk - yellow; car - red). Agents are also surrounded by a white halo when they activate the halo bias. The other options are various histograms to visualise at a glance: \textbf{values} of each mode on each criteria in the current urban setting, showing which mobility is favoured; average \textbf{priorities} of each criteria among users of each mode, illustrating the different user profiles; counters of users of each mode putting each criteria in \textbf{halo} (and thus ignoring it), allowing to spot the "weak" criterias for each mode (discrepancy between value and priority).

On the centre, the user can select a mode and a criteria in the dropdown menus, which will automatically display the up-to-date \textbf{value} of that mode on that criteria in the current town (then editable with the buttons), as well as its up-to-date \textbf{priority} in the population (average on all citizens). Under the parameters, several additional monitors also display: the average mark for that mode among its users; its average mark among users of other modes; and the list of criteria put in halo by its users (details can be seen on the halo histogram on the left). These two marks can differ because users of different modes have different priorities (which can be visualised on the priority histogram), but also because of the halo bias activated by biased citizens when evaluating their own mobility mode. 

Finally, on the right of the interface, 4 graphs display the evolution of mobility distribution (top) and average happiness (bottom) of rational (left) vs biased (right) users of each mode (one colour line per mode). This provides the user with feedback about the impact of their actions (urban planning or communication) on the mobility distribution and/or satisfaction.

\subsection{Scenarios}
Various scenarios can be tested to illustrate the potential of this simulator.

\paragraph{Urban planning to encourage soft mobility. }

We start the simulation with the default settings, then use the urban planning buttons to progressively decrease the score of car on the time criterion (meaning that the travel time by car increases). This simulates the current trend in many towns where more space is being dedicated to bicycles or buses at the expense of car facilities (driving lanes, parking spots), which increases traffic jams or time to park. The same effect could also result from a slower speed limit (30 instead of 50 km/h) that slightly increases the total trip time. 

As a result of the new setting, rational car drivers decrease their mark. They may then switch to another faster mode (such as the bus or bicycle) depending on their other priorities. We indeed observe an increase in the modal part of bus, which is closest to the priority profile of car drivers, since it preserves comfort. On the contrary, biased car drivers activate a halo on the now negative time criterion, to preserve their positive opinion of car. They continue driving to work, and their satisfaction even increases as a result of ignoring the most negative aspect.

\begin{figure}[hbt]
    \centering
    \includegraphics[scale=0.3]{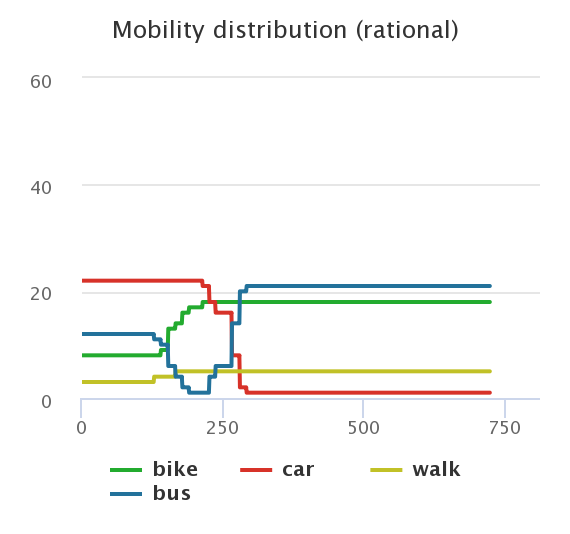}
    \caption{Rational mobility switch after ecology awareness campaign}
    \label{fig:comm1}
\end{figure}

\paragraph{Ecological crisis. }
We start the simulation with the default initialisation again, and this time use the communication buttons to increase the average priority of ecology for the population. This is meant to simulate the current energy shortages and ecological focus in European media. The resulting mobility switch is shown in Figure~\ref{fig:comm1}. Among the rational citizens, the first change happens when citizens usually commuting by bus switch to the more ecological bicycle, or to a lesser extent to walking (being much slower, it only attracts citizens with a very low priority for time). Car drivers had a lower priority for ecology at the start, so it takes longer before ecology actually impacts their decisions. When it does, they switch to the bus, which better fulfills the ecological requirement, while maintaining some level of comfort and safety (which are the next most important criteria for them). 

\begin{figure}[hbt]
    \centering
    \includegraphics[scale=0.3]{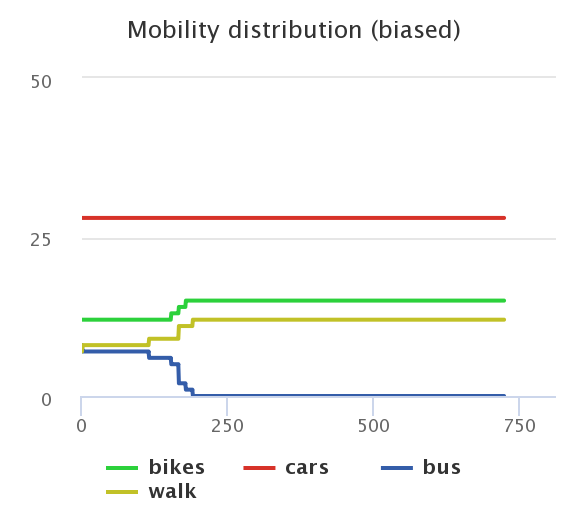}
    \caption{Biased mobility switch after ecology awareness campaign}
    \label{fig:comm2}
\end{figure}

Among biased users, things are a bit different (see Figure~\ref{fig:comm2}). Car drivers have the biggest gap between the high priority of ecology and its low value for their mode. As this gap grows bigger, it becomes more and more obvious to them that their current mode is not in agreement with their new priority. This gap grows to the point when it triggers the halo bias among all susceptible car drivers. As a result, they drop the ecology criterion in their evaluation, maintain a high satisfaction with the car, and keep driving. This might feel counter-intuitive initially, because increasing the priority of ecology leads some citizens to eventually ignore it, but it is in line with the mechanics of the halo bias. On the contrary, most users of other modes do not trigger the halo bias, because their mode mark on ecology is close enough to their priority for ecology (which was already high). Therefore, they keep using their mode if it is ecological enough, or switch to a better mode: we observe in particular that bus users switch to bike or walk.

\subsection{Discussion}

\paragraph{Realistic initial values. }
In this model, we have initialised the town and the population with believable but empirical values (initial urban planning and priorities). Real statistics could be found (for instance from INSEE for France) but it is not the goal of this work to get to that level of detail. Besides, exact figures differ for different towns or different time periods. The focus of this simulator is to illustrate the mechanism and impact of the halo bias, and as such it does not require to calibrate these values that precisely.

\paragraph{Concrete actions. }
In the scenarios above, we have investigated the impact of ``positive'' communication, trying to reinforce the priority of ecology for the population. But cars still benefit from a very positive global opinion, as conveyed by advertisement: car manufacturers often advertise their cars as related to positive values such as autonomy, safety, or adventure. It would be interesting to also study actions to restrain such (often misleading) communication.

Also, all the user's actions in the current simulator are very abstract. The simulator could be made more playful and immersive by matching the abstract actions (increasing the value or priority of a given criteria) to actual concrete actions that modify public policies or communication (\eg ``advertise about road safety'', ``raise awareness about climate change'', ``make the town centre pedestrian only'', ``make public transport free'', ``increase petrol price'', etc). However it is hard to be exhaustive when listing these actions, and several actions can have the same concrete effect on underlying values, so the current interface offers more expressive and abstraction power.

\paragraph{Cognitive biases and coping strategies. }

Our agents can use two strategies to restore cognitive consonance between their mobility and their priorities: modifying their choice of mobility, or their priorities (putting a halo on the negative criteria). This is similar to the coping strategies that are activated by individuals when they feel negative emotions. According to Lazarus \cite{lazarus1984stress}, coping strategies against negative emotions fall into two categories: either they address the stimulus itself when possible (problem-focussed coping) or they address the individual's evaluation of the stimulus when it is not controllable (emotion-focussed coping). For instance, "denial" is a strategy often triggered when someone is faced with bereavement: the negative stimulus cannot be controlled and is too intense, so the individual might react by denying its reality and refusing to believe that their relative actually died. Here agents who deal with a cognitive dissonance between their current mobility choice and their growing ecological priority might also feel that they have no control on the problem (choice of mobility), because they (believe that they) depend on their car (there might indeed be no alternative, or it would be too costly). Since they cannot or do not want to resolve the gap by changing mobility, a less costly strategy consists in denying the importance of ecology. As a result, they reappraise their current mobility without that criterion, which restores their positive opinion of it, and their satisfaction.

\section{Discussion} \label{sec:disc}

In this paper, we have presented three simulators, each focusing on a different psychological factor explaining resistance to mobility change in adaptation to the ecological crisis. All of them are available online so that people can play with them. The idea of these simulators is to explain the mechanics of these biases, so that people can become aware of their influence on their own reasoning.

\subsection{Simplification}
The goal of our simulators is not to make valid predictions about how mobility will definitely evolve in the future. Rather, they are intended to explain underlying mechanisms of some cognitive biases, and how they might influence our mobility choices. As such, we can afford to keep our simulators rather simple. In fact, since our goal is to provide interactive simulators that people can play with alone without guidance, it is even necessary to keep them simple. The relationships between inputs and outputs should stay simple to understand and visualise. 

Each simulator is thus simplified in several ways. First, contrarily to other existing models (\eg \cite{jacquier2021choice}), we did not use a real map or other geographical data (roads, buildings) from an existing town, nor did we use actual census data or mobility statistics to calibrate the population. Realistic artificial population synthesis is an active field of research \cite{chapuis2021gen}, but we do not need that level of validity here. Besides, by keeping the town abstract, we ensure that the simulators also stay generic enough. Second, each simulator only focuses on one bias in isolation, to allow visualisation of its impact without interactions with other biases that might blur the message. Of course, once again reality is much more complex. In order to confirm these simplifying modelling choices, an evaluation of our simulators will be needed on two orthogonal dimensions: validity and impact.

\subsection{Explaining human behaviour}

The first evaluation dimension is the \textbf{psychological validity} of the models, \ie how well the modelled behaviour matches real behaviour. This poses issues since it is difficult to get data about how people really think and make decisions. Biases are not conscious, so what people report in surveys does not necessarily match what really led their decisions. Besides, our interpretation of such data would be subjective, \ie even if the simulated behaviour matches our data, it does not mean that the underlying model is correct. Indeed, several different models can lead to the same observed behaviour. 

Indeed, researchers have identified a number of cognitive biases in interviews conducted with survivors of the Australian Black Saturday bushfires in 2009 \cite{adam2017modelling}, that could explain their decision not to evacuate; but the authors also showed how several biases can explain the same behaviour. Other researchers have analysed the conversation between the pilot and copilot from the flight recorder (black box) after a plane crash \cite{crash1982} and hypothesized that the self-deception bias could explain this crash; but again this is subject to interpretation, and other explanations could be proposed. Recent work \cite{fouillard2021catching} even proposed a logical approach to automatically deducing which biases could be involved in an erroneous decision process, but again several different explanations can be inferred from the same incident. 

This is the case even with the three simulators proposed here. They can all provide an explanation to the inertia observed in people's adaptation to climate change: people do not change mobility because their habits bypass rational evaluation; or they do not change mobility because the halo bias keeps them satisfied with their current mode; or they do not change mobility because they want to assert their free will to do as they please. All explanations work, and are probably correct for different individuals, depending on their personality and susceptibility to one or the other bias. Further, many other explanations could probably still be found. This reasserts our goal, which is to raise awareness in people and give them food for thought, rather than provide a definite explanation.

\subsection{Impact on debiasing}
The second dimension of evaluation is the \textbf{impact} of these simulators on people. We claimed that they could be used as a debiasing intervention, but an evaluation will be necessary to prove this claim. This is easier to do: we plan to run an online survey where people will be interrogated before and after playing the simulator. We are also considering logging their actions in the simulator to directly analyse how they adapt their strategy when they realise the impact of the bias on the simulated population. Finally, we could compare the impact of various types of interventions, such as interacting with our simulator, reading a text explaining the bias, or watching an explanatory video.

It is important to reaffirm here that we do not intend to directly change people's behaviour. It is well-known in serious games research that debriefing is an essential part of learning through a game \cite{crookall2010serious,whalen2018all,dieleman2006games}. In the current setting where our simulators are played online in autonomy, we cannot therefore expect much learning to happen. We do however expect that it will make people reflect about their mobility and their cognitive biases. This reflection might participate in a change over the longer-term, but this is hard to track and prove anyway.

\subsection{Equity}

Besides, it would be overly optimistic to expect that just becoming aware of potential cognitive biases in mobility choices will suffice to raise the modal part of soft mobilities such as bicycle or walking. In reality, individual constraints (reduced physical fitness, family commitments \eg picking kids from school, tight time schedule or limited budget) as well as the urban setting (lack of public transport in many rural areas, concentration of bicycling facilities in town centres) also play an essential role in mobility choices. 

Actually, many public policies aimed at changing mobility choices tend to forget those who cannot afford such a change. For instance low emission zones will restrict access to oldest cars, but some people cannot afford buying a brand new electric car. People living further from the town centre already spend a lot of time commuting, and cannot afford to spend even more time to commute by bike or bus. Elderly people cannot move very far by foot, so they depend on their car or public transport: pedestrianising town centres might exclude them unless shuttles ensure access. The list of criteria is long still. To address this question of equity, in other work we are designing an interactive simulator where the user tries to modify the urban infrastructures dedicated to the different mobilities, in order to raise the modal part of soft mobility, but also while preserving a number of equity indicators, for instance accessibility computed as the percentage of citizens who do have an available mobility mode.

\section{Conclusion} \label{sec:cci}
\enlargethispage{15pt}

This paper introduced three interactive simulators, each focusing on the impact of one particular cognitive bias on mobility inertia in the face of climate change and evolving urban planning. All simulators are available online. Our goal is not to claim any definitive explanation to mobility inertia, but rather to let users discover and visualise the impact of biases on mobility. We want to give them food for thought. 

Even though we have illustrated in this paper the explanatory power of each simulator on a number of scenarios, we have not yet conducted a real evaluation. Short-term future work will be dedicated to evaluating the actual impact of each simulator on users via an online survey. On the longer-term, we will study the integration of these biases in \anonymous{a more complex serious game}{the SWITCH serious game \cite{gamadays22}}. In this game, the player takes on the role of the urban manager, who must modify the urban planning in their town in order to adapt to climate change: they face contradictory goals of reducing pollution and sedentarity, while maintaining accessibility and satisfaction. To make the game more realistic, the population has already been endowed with habits. Future work will also integrate the impact of cognitive biases, including those modelled here (reactance and halo) as well as other relevant biases (social pressure for instance). The goal of this game will be to support prospective thinking about what the mobility of tomorrow could look like if we overcome our resistances to change. Imagining a better future could be the first step before making it a reality.

\end{document}